\documentclass[sigconf,authorversion]{acmart}

\AtBeginDocument{%
  \providecommand\BibTeX{{%
    \normalfont B\kern-0.5em{\scshape i\kern-0.25em b}\kern-0.8em\TeX}}}

\copyrightyear{2020}
\acmYear{2020}
\setcopyright{acmlicensed}
\acmConference[ACSAC 2020]{Annual Computer Security Applications Conference}{December 7--11, 2020}{Austin, USA}
\acmBooktitle{Annual Computer Security Applications Conference (ACSAC 2020), December 7--11, 2020, Austin, USA}
\acmPrice{15.00}
\acmDOI{10.1145/3427228.3427233} \acmISBN{978-1-4503-8858-0/20/12}

\usepackage{multirow,color}  
\catcode`\_=12  
\graphicspath{{./img/}}

\usepackage{graphicx,tikz}

\newcommand*\encircle[1]{\tikz[baseline=(char.base)]{
            \node[thick,shape=circle,fill,inner sep=1pt,font=\sffamily] (char) {\textcolor{white}{#1}};}}

\newcommand{\BLINDED}[1]{#1} 

\newcommand{\enquote}[1]{`#1'}  

\newcommand{\arch}[1]{\texttt{AC-#1}}
\newcommand{\require}[1]{\texttt{R-#1}}

\newcommand{\SOC}[0]{SOC}

\newcommand{\GOODSOC}[0]{\texttt{GOODSOC}}
\newcommand{\BADSOC}[0]{\texttt{BADSOC}}

\newcommand{\SAIBERSOC}[0]{\texttt{SAIBER\-SOC}}

\newcommand{\MIRAI}[0]{\texttt{Mirai}}
\newcommand{\EXIM}[0]{\texttt{Exim}}

\newcommand{\MIRAIATTACKERIP}[0]{199.19.215.23}
\newcommand{\EXIMATTACKERIP}[0]{31.220.56.38}

\newcommand{\MIRAIVICTIMIP}[0]{131.155.68.116}
\newcommand{\EXIMVICTIMIP}[0]{131.155.71.27}

\acmBadgeR[https://www.acsac.org/2020/submissions/papers/artifacts/]{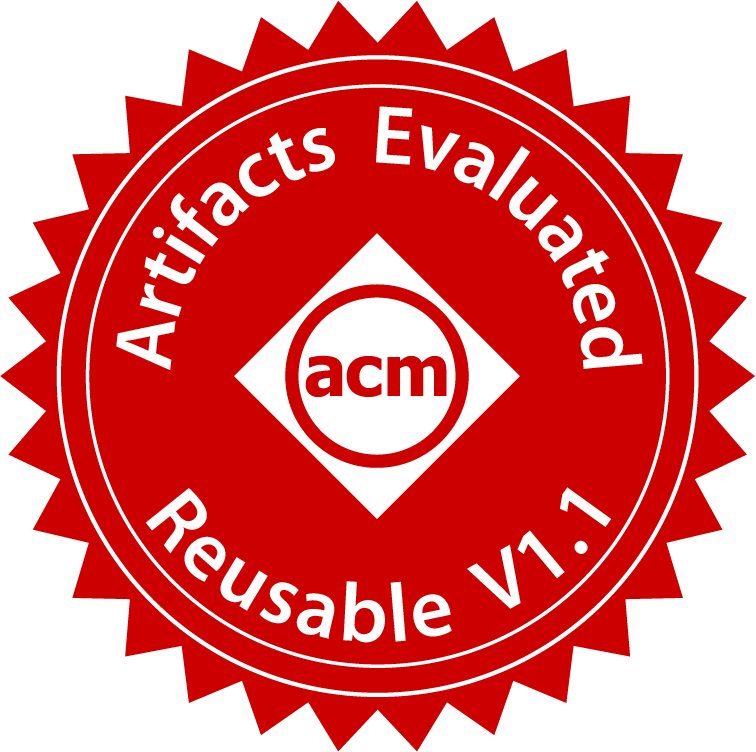}

\begin{document}

\title{\SAIBERSOC{}: Synthetic Attack Injection to Benchmark and Evaluate the Performance of Security Operation Centers}

\author{Martin Rosso}
\email{m.j.rosso@tue.nl}
\affiliation{%
  \institution{Eindhoven University of Technology}
  \streetaddress{P.O. 5600 MB Eindhoven}
  \city{Eindhoven}
  \country{The Netherlands}}
\author{Michele Campobasso}
\email{m.campobasso@tue.nl}
\affiliation{%
  \institution{Eindhoven University of Technology}
  \streetaddress{P.O. 5600 MB Eindhoven}
  \city{Eindhoven}
  \country{The Netherlands}}
\author{Ganduulga Gankhuyag}
\email{g.gankhuyag@tue.nl}
\affiliation{%
  \institution{Eindhoven University of Technology}
  \streetaddress{P.O. 5600 MB Eindhoven}
  \city{Eindhoven}
  \country{The Netherlands}}
\author{Luca Allodi}
\email{l.allodi@tue.nl}
\affiliation{%
  \institution{Eindhoven University of Technology}
  \streetaddress{P.O. 5600 MB Eindhoven}
  \city{Eindhoven}
  \country{The Netherlands}}

\renewcommand{\shortauthors}{M. Rosso, M. Campobasso, G. Gankhuyag, and L. Allodi} 

\begin{abstract}
  In this paper we introduce \SAIBERSOC{}, a tool and methodology enabling security researchers and operators to evaluate the performance of deployed and operational 
  Security Operation Centers (SOCs) (or any other security monitoring infrastructure). The methodology relies on
  the MITRE ATT\&CK Framework to define a procedure to generate and
  automatically inject synthetic attacks in an operational SOC to evaluate any
  output metric of interest (e.g., detection accuracy, time-to-investigation,
  etc.). To evaluate the effectiveness of the proposed methodology, we devise
  an experiment with $n=124$ students playing the role of SOC analysts. The
  experiment relies on a real SOC infrastructure and assigns students to
  either a \BADSOC\ or a \GOODSOC\ experimental condition. Our results show
  that the proposed methodology is effective in identifying variations in SOC
  performance caused by (minimal) changes in SOC configuration.
  We release the \SAIBERSOC{} tool implementation as free and open source software.
\end{abstract}



\maketitle


\section{Introduction}

The growing importance of effective security monitoring solutions calls for appropriate measures of their effectiveness.
The operation of Security Operation Centers (SOCs) is the recommended best practice to which large and medium-size enterprises rely for the detection, notification, and ultimately response to cybersecurity incidents~\cite{kokulu2019matched,muniz2015security}.
Yet, the average time for detecting an attack ranges between several weeks to years~\cite{verizon2018}. In a recent study, Kokulu et al.~\cite{kokulu2019matched} interviewed security analysts and mangers, who explicitly identify the issue of \textit{measuring} SOC performance as one of the main obstacles towards effective detection and response operations.

The problem of measuring security (performance) is a long-standing and difficult one~\cite{herley2017sok}.

Measuring the performance of \SOC{}s is no exception.
Whereas numerous performance metrics exist (e.g., number of detected incidents, time to detection, time to response, etc.)~\cite{shah2018understanding}, it is still unclear how to obtain reliable and reproducible measures over those metrics.
As incoming attacks are, by definition, unknown, a ground truth on which to base the measurements cannot be easily defined;
on the other hand, performing measurements in \textit{in-vitro} settings limits the representativeness of those measurements in real-world settings~\cite{shah2018methodology,kokulu2019matched}.

However, the problem of measuring the performance of infrastructures and systems dedicated to the detection of \textit{rare} events of unknown magnitude is not new.
For example, the \textit{LIGO} (\textit{Laser Interferometer Gravitational-Wave Observatory}) infrastructure has been built with the goal of detecting perturbations in the spacetime continuum caused by `rippling effects' of large-scale events, such as the collapse of a binary blackhole system. Despite the completely different settings and event generation processes, the challenges faced by the LIGO interferometer and SOCs are quite similar: both must detect unknown, unpredictable, and arbitrarily `small' manifestations of an event, and both must know whether the detection procedure is accurate while operating in the absence of a ground-truth on the events that ought be detected.

This paper presents \texttt{SAIBERSOC}, a practical and deployable solution for SOC performance evaluation based on the same working principle as the LIGO infrastructure~\cite{PhysRevD.81.102001}: to generate and automatically inject `events' (in our case, cyber-attacks) into the detection infrastructure at random instants in time, with the purpose of evaluating, in a controlled fashion, the performance of the detection system.
The \SAIBERSOC{} methodology and tool can be employed by researchers and security operatives alike to evaluate the impact of various factors on SOC performance, such as analyst training, experience , and skill composition, or the relative change in effectiveness of the deployed analysis and visualization tools as the SOC configuration and processes change.

\SAIBERSOC{} automatically generates the injected attacks relying on the MITRE ATT\&CK Framework\cite{mitre_attck} and is structured over four components providing automated methods for attack definition, generation, injection, and analysis.
The \SAIBERSOC{} tool is publicly available as free and open source software (see subsection~\ref{subsec:saibersoc-tool}).

To evaluate the effectiveness of the method at the core of \SAIBERSOC{}, we design and run an experiment employing $n=124$ students enrolled in a security course of a medium-sized technical university in Europe.
The experiment relies on a real SOC infrastructure (normally monitoring network events at the M\&CS department of the university) where students operate as SOC analysts.
In particular, we measure detection accuracy in terms of both attack \textit{identification} and \textit{investigation}, injecting two attacks based on the \MIRAI{}~\cite{mirai-botnet} and \EXIM{}~\cite{exim-cve} scenarios over two SOC configurations (\BADSOC{} and \GOODSOC{}).
Our evaluation shows that the \SAIBERSOC{} methodology correctly identifies and can quantify the relative change in detection accuracy across SOC configurations and, importantly, does not \enquote{overshoot} or reveal differences where there should be none.

\smallskip
This paper develops as follows: \hyperref[sec:background]{Section~\ref{sec:background}} provides a background on SOC operations and discusses relevant literature.
\hyperref[sec:method]{Section~\ref{sec:method}} describes the proposed methodology and section~\ref{sec:validationmethodology} the experimental validation strategy.
Experimental results follow in section~\ref{sec:experimentresults}.
\hyperref[sec:discussion]{Section~\ref{sec:discussion}} discusses our results and presents the \SAIBERSOC{} software implementation in detail.
Finally, section~\ref{sec:conclusion} concludes the paper.

\section{Background and Related Work}
\label{sec:background}

\subsection{Attack frameworks}
Defensive operations against increasingly sophisticated attacks require a deep understanding of how adversaries conduct offensive operations, how threats unfold and how security breaches escalate into major incidents.
The MITRE PRE-ATT\&CK\cite{mitre_pre-attck}, MITRE ATT\&CK \cite{mitre_attck}, and Cyber Kill Chain\cite{lockheed_martin} frameworks aim at building a knowledge base of adversary TTPs (Tactics, Techniques, and Procedures) from real-world observations, grouping them in a taxonomy of attack stages to describe the anatomy of any attack.
The extensive and detailed enumeration of the attack techniques provided by the frameworks and stages can be, roughly, grouped in four phases: \textit{Reconnaissance}, \textit{Exploitation}, \textit{Delivery} and \textit{Control} (Table~\ref{tab:phases}).
\textit{Reconnaissance} groups data gathering operations performed by the adversary, ranging from subject information for the development of social engineering attacks, to probing the network surface of the infrastructure to derive paths of compromise. 
During \textit{Exploitation}, the attacker evaluates the measured attack surface (vulnerabilities, misconfigured software, phishing opportunities) and creates or acquires the attacks (vulnerability exploits, phishing artefacts, malware, ...) to perform information leak, lateral movement (e.g., inside the organization), command execution, and/or privilege escalation.
\textit{Delivery} covers the vectors adopted from the adversary to drop malicious software that will act as a foothold for further actions. This includes spear phishing and supply chain compromise.
\textit{Control} includes the techniques adopted by the adversary to control the system and reliably execute commands on it, compromising system integrity, confidentiality and availability. This can be achieved via remote management interfaces, custom software and web services. 

\begin{table}[t]
    \centering
    \caption{Attack phases}
    \label{tab:phases}
    \Description{Overview of the four phases `Reconnaissance', `Exploitation', `Delivery', and `Control'}
    \begin{tabular}{p{0.22\columnwidth} p{0.7\columnwidth}}
        \toprule
        \textbf{Phase} & \textbf{Description}                                                                            \\\midrule
        Reconnaissance & Techniques to research, identify and select targets using active or passive reconnaissance.     \\
        Exploitation   & Techniques employed by attackers to gain initial control over (vulnerable) target systems.      \\
        Delivery       & Techniques resulting in the transmission of the weaponized object to the targeted environment.  \\
        Control        & Techniques used by attackers to communicate with controlled systems in a target network.        \\\bottomrule
    \end{tabular}
\end{table}

\subsection{Security monitoring operations}
Tools and procedure to support security analysts in incident response and network monitoring (such as Network Intrusion Detection Systems (NIDS) and Security Information and Event Management (SIEM)) are at the core of modern security monitoring in operational settings. \textit{Security Operation Centers} (SOCs) are the center of monitoring operations in medium or large organizations, either internally or outsourced to service providers~\cite{muniz2015security,kokulu2019matched}. 

SOCs are organized hierarchically in (generally three) tiers~\cite{muniz2015security}, where analysts with different skills and expertise monitor the network activity and take action against a threat.
Tier~1 is the first frontier where alerts are investigated by analysts, identifying possible threats among the non-significant ones and prioritizing them.
Identified threats are escalated to Tier~2, where more qualified analysts with forensics and incident response skills correlate the information with threat intelligence to identify threat actors.
Tier~2 is in charge to determine a strategy for containment, remediation and recovery.
If the threat targets business critical operations, Tier~3 analysts identify and develop tailored responses to the identified threats and attack patterns.

In general, incident investigation can be split in two phases~\cite{muniz2015security}: \textit{Attack identification}, where Tier 1 analysts evaluate incoming alerts to identify possible attacks; \textit{Attack investigation}, where Tier~1 and Tier~2 analysts investigate the identified attack over its phases (ref. Table~\ref{tab:phases}) to, for example, identify victims, attack timing, payloads, and propagation. Once these two phases are complete, the attack is reported and responded to, depending on the service level agreement at which the SOC operates.

The large amount of alerts and potential security events detected in a SOCs make it impossible, operationally to `investigate everything'~\cite{shah2018understanding}: the more relevant are the alerts generated by a SOC (e.g., through accurate use-cases), the fewer `false positives' a (Tier~1) analyst will have to investigate before passing the baton on to higher tiers (i.e., Tier~2~and~3).

However, \textit{measuring} how SOC operations respond to changes in the SOC configuration (e.g., a refined use-case, a different alert investigation process, new rulesets, etc.) remains an open and critical challenge~\cite{kokulu2019matched}, in terms of analyst competences as well as for metric definition and measurement \cite{shah2018understanding,shah2018methodology,zimmerman2014ten,ganesan2017optimal}.

\subsection{SOC performance evaluation}
Albeit attempts have been made in trying to tackle these problems, the scientific literature proposing methods to measure SOC performance is still limited.
Effectiveness of a SOC depends on both the qualities of the tools adopted, their configurations and capabilities, and on skills of the personnel. For what concerns the correlation between human capabilities and a SOC's performance, a few studies have been made\cite{sundaramurthy2016turning, sundaramurthy2014tale}. In \cite{sundaramurthy2014tale}, Sundaramurthy et al., they trained three students to work as SOC analysts and embedded them in three different SOCs. While working as operational components, they were required to produce reports on their observations and meet with an anthropologist involved in the research. By comparing reports and interviews, is was possible to draw conclusions on the aspects that may be more promising to the effectiveness of a SOC and which not, in terms of its architecture, experience of analysts, nature and degree of their interactions, workflow for incident reports and the quality of work itself for the employee. 
Kokulu et al.~\cite{kokulu2019matched} investigate through interviews to both SOC analysts and SOC managers about technical and non-technical issues of a SOC; from their interviews emerge how often SOCs fail to provide substantial support against specific types of attacks, overloading analysts with low-quality threat intelligence, long reports and logs, and more. Albeit evidences are supported by interviews, the authors point out that metrics to measure security quantitatively and qualitatively are rudimentary \cite{kokulu2019matched}.
Jacobs et al.~\cite{jacobs_towards_2014} propose a method to systematically and quantitatively evaluate \SOC{}'s maturity by identifying a set of capabilities that a \SOC{} embodies, i.e. log analysis, event correlation, incident management, threat identification and reporting, scoring each of them \cite{jacobs_classification_2013} \cite{jacobs_towards_2014}. The SOC aspects analyzed derive from a number of industrial security management and control frameworks, including ISO 27000 series \cite{iso_2019} and SANS Critical Controls and each of them is scored with respect to their maturity;
however, the score attribution is left as an expert-driven task, rather than a data-driven measurement.
A simulation-optimization approach is proposed by Shah et al. \cite{shah2018methodology}, where they identified some of the causes that negatively affect the throughput of a SOC in terms of efficiency. They propose the \textit{Time to Analyze Alert} (TTA) metric to evaluate the efficiency of a SOC by measuring the time that goes from the alert generation to its analysis. By monitoring SOC's TTA, Shah et al. are able to generate advice for the SOC Manager, suggesting live corrective actions towards a desired benchmark. In subsequent work Shah et al.\cite{shah2018understanding} also model the problem as a simulation-optimization problem and evaluated it from data derived from a simulated SOC, to obtain a set of metrics that allow the optimization of some of a SOC capabilities; however, a procedure to test the resulting SOC performance in real-world scenarios remains to be identified and tested.

\smallskip
\textbf{Research gap.} Whereas current research has focused mostly on the identification of metrics~\cite{shah2018understanding} and procedures~\cite{jacobs_towards_2014,shah2018methodology,shah2018understanding} to evaluate SOC performance, an empirical method capable of capturing the complexity of a SOC operation (including alert configuration, analyst capabilities, etc.) has yet to be proposed and validated~\cite{kokulu2019matched}.

%
\section{The \SAIBERSOC{} solution}
\label{sec:method}

To address this gap, we propose \SAIBERSOC{}, a method and solution to perform systematic SOC performance evaluations through automated attack injection. Our solution is composed of four Architectural Components (\texttt{AC-\{1..4\}}) constituting of a library of (attack) traces (\arch{1}), and components to generate (\arch{2)}, inject (\arch{3}), and report (\arch{4}) the synthetic attacks.

\begin{figure}[t]
    \centering
    \includegraphics[width=0.73\columnwidth]{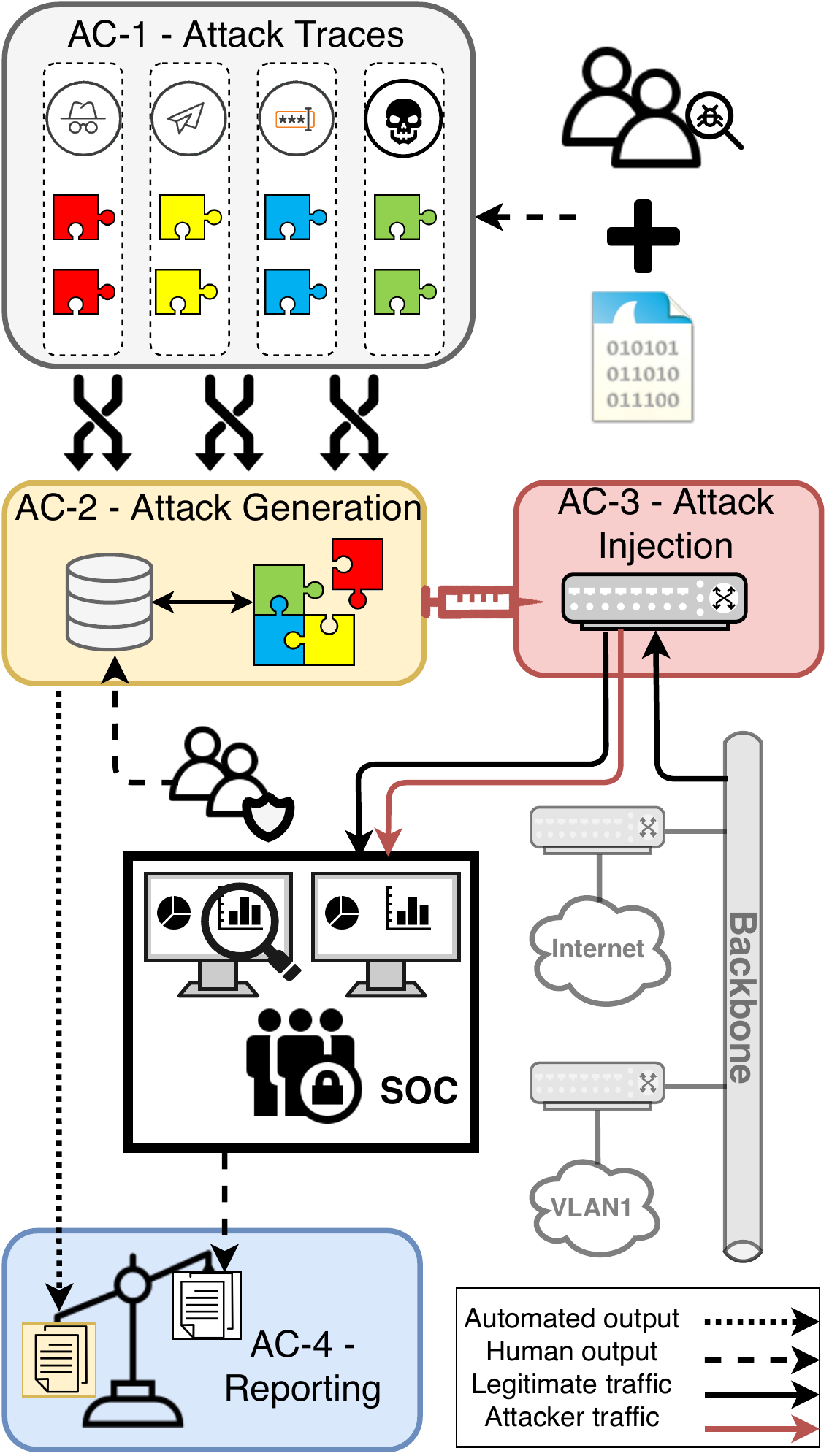}
    \begin{minipage}{0.9\columnwidth}
        \footnotesize
        \smallskip
        The platform uses (network) traces characterizing the \enquote{building blocks} of an attack (\arch{1}), and (re-)combines them to generate new attacks (\arch{2}).
        The generated attacks are then injected into the network (\arch{3}).
        The SOC analysts perform and report on the investigation results, which can be checked (\arch{4}) against the known ground-truth (derived from the  attacks generated in \arch{2}).
    \end{minipage}
    \caption{Schematic depiction of the \SAIBERSOC{} solution}
    \label{fig:schema}
    \Description[Schematic overview of how the four components interact in relation to the Security Operations Center.]{Component `AC-1 Attack Traces' is a library of network traffic captures maintained by experts. In component `AC-2 Attack Generation' the network traces get merged. Experts monitor and supervise rules for merging network traces to ensure that the output is a realistic attack. The output of this component is the assembled attack ready to be replayed on the network, as well as a ground-truth that describes the assembled attack. In phase `AC-3 Attack Injection' the assembled attack is injected somewhere into a network monitored by the Security Operations Center. The Security Operations Center, unable to distinguish injected attack simulation traffic from regular network traffic, will investigate and report on the injected attack. In component `AC-4 Reporting' the report generated by an analyst in the Security Operations Center is compared to the automatically generated ground truth from `AC-2 Attack Generation'.}
\end{figure}

\SAIBERSOC{} can be used to inject attacks into a \SOC{} during operation, or alternatively in a \enquote{virtual} \SOC{} testing environment where all network traffic is replaced by pre-recorded background traffic.
Once the attack has been detected, the \SOC{} analysts perform the investigation following their normal procedure.
The type of injected attack as well as its timing, targets and sources, and velocity are unknown to the \SOC{} analysts.
Finally, the analyst's attack report is automatically

evaluated against the ground truth obtained from the attack assembly phase.
A more detailed description of the \SAIBERSOC{} architecture can be found in subsection~\ref{subsec:architecture}, we present our software implementation in subsection~\ref{subsec:saibersoc-tool}.

%


\subsection{Solution requirements}

To guide the development of the proposed solution, we pose a list of requirements that a real-world \SOC{} performance evaluation framework should fulfill. A summary of the identified requirements is provided in Table~\ref{tab:requirements}.
\begin{table*}[t]
    \centering
    \caption{Architectural requirements. Table~7 in the Appendix details the mapping to the architectural components.}
    \label{tab:requirements}
    \Description[Mapping between requirements `R' and architectural components `AC' that must meet the respective requirement.]{Architectural components `AC-1', `AC-2', and `AC-4' must fulfill requirement `R-1 Independence' (the proposed solution should be independent of the specific SOC implementation and monitored environment). Architectural components `AC-1' and `AC-2' must fulfill requirement `R-2 Versatility' (the framework must work with all types of attack scenarios in scope of the SOC, regardless of their complexity). Architectural components `AC-2' and `AC-3' must fulfill requirement `R-3 Realism' (the attack injection must be seamlessly integrated to the operative environment of the analyst. This includes IP addresses, local software configurations, and network setups). Architectural components `AC-2' and `AC-4' must fulfill requirement `R-4 Tracability' (the framework must be able to link all SOC reactions i.e., a report to the initial stimulus or attack that caused it.).}
    \begin{tabular}{l l p{0.56\textwidth} c c c c}
    \toprule
         \require{ID} & \textbf{Requirement} & \textbf{Description} & \arch{1} & \arch{2} &\arch{3} &\arch{4} \\
         \midrule
         \require{1} & Independence & The proposed solution should be independent of the specific SOC implementation and monitored environment.                                                                 & \checkmark & \checkmark &            & \checkmark \\
         \require{2} & Versatility & The framework must work with all types of attack scenarios in scope of the \SOC{}, regardless of their complexity.                                                         & \checkmark & \checkmark &            &            \\
         \require{3} & Realism & The attack injection must be seamlessly integrated to the operative environment of the analyst. This includes IP addresses, local software configurations, and network setups. &            & \checkmark & \checkmark &            \\
         \require{4} & Traceability & The framework must be able to link all \SOC{} reactions (i.e., a report) to the initial stimulus or attack that caused it.                                                &            & \checkmark &            & \checkmark \\
         \bottomrule
    \end{tabular}
\end{table*}

\noindent\textbf{Independence (\require{1})}
To measure and compare \SOC{} performance, the framework must apply to all \SOC{}s, regardless of how they are implemented.
Further, the results must be reproducible, regardless of who uses the framework.
Those are fundamental requirements for any measure or assessment~\cite{pfleeger2010measuring}.
Therefore, a solution must be able to work with various \SOC{} configurations and implementations.

\noindent\textbf{Versatility (\require{2})}
To be able to create meaningful measures covering the whole scope of the \SOC{}, all aspects of a potential attack must be covered.
Thus, a metric should cover the complete value-space to assure  meaningful results reflecting real-world performance~\cite{kokulu2019matched}.
Because attackers continuously adapt their behavior, a solution must be capable to simulate past, current, as well as future attacks in realistic ways to be able to measure real-world security.

\noindent\textbf{Realism (\require{3})}
Real-world security is influenced by numerous factors, sometimes in unpredictable ways.
For example, wrong presumptions about context or human behavior in unforeseen situations leads to alleged security measures and therefore a false sense of security\cite{chiasson_quantifying_2015,adams2005bridging}.
This can be avoided by assessing real-world performance, i.e., a setup that is as close as possible) to reality.
In case of a \SOC{} this means measuring a real \SOC{} (including software and configuration), realistic work environments, and realistic attacks. 
It is therefore imperative that the evaluation is performed under real or realistic operational conditions (e.g., the analysts do not know the attacks in advance, and the injected attacks are realistic).

\noindent\textbf{Traceability (\require{4})}
The framework must be able to link \SOC{} reactions i.e., reports, to the attacks that caused those reactions in the \SOC{}.
Only if both input and output are known, a meaningful evaluation of the \SOC{} performance can be made by comparing the \textit{expected} output to the realized output.

%
\subsection{The \SAIBERSOC{} solution architecture}
\label{subsec:architecture}

Figure~\ref{fig:schema} provides a schematic depiction of the architecture behind the proposed \SAIBERSOC{} solution.
Table~\ref{tab:requirements} reports the mapping of the architectural components to the identified requirements. 
A running example of implementation of the proposed workflow is reported in Table~\ref{tab:example} in the appendix and discussed below.

\subsubsection{Attack Traces (\arch{1})}
\label{subsubsec:attack-module-database}
All attack simulations or scenarios deployed in \SAIBERSOC{} are constructed by combining multiple pre-defined attack traces. An attack trace consists of a set of network packets (e.g., stored in a \textit{pcap} file) that reproduces the network traffic generated by an attack during a specific attack phase.
These traces can be extended to non-network events (e.g., \textit{syslog} activity to reproduce events at the host level).
To structure traces in attack phases, we rely on the ATT\&CK Framework\cite{mitre_attck} to create and store a library of different attack traces reflecting different phases of ongoing attacks (see Table~\ref{tab:phases}).

The MITRE CALDERA\cite{caldera} and Atomic Red Team\cite{atomic} already maintain some form of an attack-database with attacker actions mapped to the ATT\&CK Framework\cite{mitre_attck}.

\paragraph{Impl. example:} with reference to the example reported in Table~\ref{tab:example}, \arch{1} generates a set of traces characterizing events in each phase of the ATT\&CK framework; for example, \textit{portscan(src\_ip, dst\_ip)} will reproduce network traffic (to be parameterized with \textit{src\_ip, dst\_ip} associated to the \textit{Reconnaissance} phase of the MITRE ATT\&CK framework (ref. Table~\ref{tab:phases}). Similarly, \textit{expl\_cve()} will reproduce traces corresponding to the exploitation of a known vulnerability (e.g., as derived from an exploit on ExploitDB~\cite{exploitdb}.

\subsubsection{Attack Generation (\arch{2})}
\label{subsubsec:scenario-builder}

Multiple attack traces are combined to generate fully-fledged attack simulations.
Attacks are composed following the ATT\&CK framework\cite{mitre_attck} (i.e., reconnaissance happens before exploit delivery, which happens before command and control communication, which happens before data exfiltration).

The CALDERA framework is already capable of automatically generating meaningful attacks from attack traces\cite{caldera}.

In this step, the generic attack traces are matched to the specific environment monitored by the SOC (e.g., IP ranges, exploits vs vulnerable systems, etc.). This process can be automated by relying on asset information available to any SOC during normal operation.

The result of this phase is a \textit{ready-to-inject} attack simulation.
 
The \enquote{ground-truth} of the simulation (e.g., attacker's IP addresses(es)) can be extracted from the parameterized attack traces composing the attack. 
Once re-parameterized (i.e., victim IP addresses matching the range of a monitored network and system functionality), attack scenarios can be reused the \SOC{} or organization's network.

\paragraph{Impl. example:} with reference to Table~\ref{tab:example},
a sequence of (attack) traces is selected and parameterized.
In this example, the IP of the attacker and the receiving victim system(s)  are identified (\textit{src\_ip,dst\_ip}), as well as the exploited vulnerability (\textit{cve}), and protocols used for the command and control (\textit{proto}).
The parametrized attack traces are then composed following the MITRE ATT\&CK framework phases; the parameters of the attack are used to generate the ground truth against which the analysts' reports will be checked in \arch{4}.

\subsubsection{Attack Injection (\arch{3})}
\label{subsubsec:attack-scenario-injector}

Attack simulations retrieved from \arch{2} are injected in the network traffic monitored by the \SOC{}.
This can be either executed on the live infrastructure, or on a virtual \enquote{shadow network} separate from the operative environment.
In the latter setting, the entire (virtual) network traffic is recorded and later forwarded to the Network Intrusion Detection sensors employed by the \SOC{}.
The attack simulation can be merged seamlessly into (real-world) traffic streams either on a \SOC{} sensor or on network infrastructure devices (i.e., network switches or routers).
To make simulations as realistic as possible, (optional) background traffic that was previously recorded on site can be mixed with the attack simulation.
Attack simulations can be injected, recorded, and replayed at any point in time (i.e., record network traffic from virtual environment instead of passing to the \SOC{} immediately).
The \SOC{} will then raise the respective events and alerts as if the injected traffic was real traffic to (or from) the monitored infrastructure. 

\paragraph{Impl. example:}
In the example in Table~\ref{tab:example} traffic is recorded and replayed into a \SOC{} sensor as scheduled by executing the parametrized traces (e.g., \textit{exploit\_cve(cve,src\_ip,dst\_ip)}) (i.e., 12pm).
The attack injection raises corresponding \SOC{} alerts, then evaluated by the analyst during incident identification and investigation.

\subsubsection{Reporting (\arch{4})}
\label{subsubsec:report-evaluator}

The attack reports produced by \SOC{} analysts in response to the attack simulation are matched against the ground truth (obtained from \arch{2}) in \arch{4}.
Because the exact details of the simulated attacks are known in advance, report(s)  that mention specific details unique to the injected attack can be identified automatically.
Using simple string matching and the knowledge-base of the SIEM, it is possible to check whether the final \SOC{} report details the incident events and sources correctly.

\paragraph{Impl. example:}
A \SOC{} analyst investigated the generated SOC alerts and reported the incident through a standardized template (e.g., \textit{src\_ip, cve}, \ldots).
Comparing the analyst report with the ground truth may indicate that the current \SOC{} configuration lacks detection capabilities (e.g., undetected  exploitation), or that raised alerts systematically misdirect the analyst (e.g., due to poorly defined threats or use cases).

%
\section{Experimental validation}
\label{sec:validationmethodology}

Before presenting the \SAIBERSOC{} tool implementing the described architecture and solution, we validate the methodology behind it by testing it experimentally. To do so, we devise an experiment involving $n=124$ MSc students attending the 'Security in Organizations' course of a medium-sized European Technical University and an operative SOC infrastructure deployed at the Department level.

\smallskip
\textbf{Goal of the experiment}: \textit{to evaluate whether the proposed solution is capable of detecting (small) differences in SOC performance.}

\smallskip
Experimental subjects worked in groups of two and took the role of \textit{Tier 1} security analysts to investigate (injected) attacks.
Subjects were randomly assigned to the experimental treatment (consisting in a small tweak in the SOC configuration) and asked to report the results of their analysis by filling in a report template.

\subsection{Experimental infrastructure}

The experiment is based on a real SOC used for research and education operating at the Mathematics and Computer Science department of the university. The SOC is operated over open-source technology and based on Security Onion\cite{onion} and Elastic Stack\cite{elastic} for event correlation and analysis. During regular operation the SOC monitors real incoming and outgoing network traffic in the department. The chosen network sensors (NIDS) are Suricata\cite{suricata} for security monitoring, and Bro/Zeek\cite{zeek} for network flow logs.
The SOC is currently limited to the monitoring of network traffic and is operated with the involvement and cooperation of the security team of the university.

\subsection{Experimental subjects}

Subjects were recruited from a security in organization MSc course held at the university operating the SOC. The course is mandatory for all students graduating in the security track of the MSc program in Computer Science and open to students from others tracks (security=14; computer science=56; other=64; tot=124).
Students were asked to form groups of two to participate in the experiment and were randomly assigned to the respective treatment groups (next subsection).
The final size of the experiment pool is n=63; the term `experiment subject' in the remainder of the paper will refer to the student groups and not to the single students.

\subsection{Experimental variables}
\label{subsec:experimental-variables}

\paragraph{Independent variable} To reproduce realistic SOC configurations we setup two analysis environments, namely \GOODSOC{} and \BADSOC{}. To test the effectiveness of the proposed solution, we introduce only a small change between the two configurations. To maintain the change realistic, we (1) define it around the university environment in which the SOC operates, and (2) base it only on the (de-)activation of a set of predefined rules that \textit{do not} match the use-cases of the university.
\footnote{This setup has been checked against a set of use-cases provided by the university for SOC operation.}
Namely, as the university is an open environment, default rulesets triggering alerts related to violation of generic policies such as use of TOR or p2p are out of the scope of the SOC. We therefore define the following two experimental conditions:

\smallskip
\noindent\textbf{\BADSOC{}}: baseline SOC configuration consisting of the default set of rules defined by the detection software (Suricata);\hfill\break
\textbf{\GOODSOC{}}: baseline SOC configuration \textit{minus} alerts related to policy violation events.

\smallskip
\noindent In the \BADSOC{} configuration 19731 out of 27125 present Suricata~\cite{suricata} rules were active by default.
By deactivating 2753 rules (14\%), 16978 active rules remain).
The disabled rules\footnote{All rules prefixed with either \enquote{ET POLICY}, \enquote{ET INFO}, \enquote{ET CHAT}, \enquote{GPL CHAT}, \enquote{ET TOR}, or \enquote{SURICATA} were disabled (see \enquote{\href{https://gitlab.tue.nl/saibersoc/acsac2020-artifacts/-/blob/master/experiment/scripts/better-soc.sh}{\texttt{better-soc.sh}}} in the \href{https://gitlab.tue.nl/saibersoc/acsac2020-artifacts/}{artifact repository}).} inform about policy violations (including the use of chat/instant messaging software and usage of the TOR network\cite{tor}).

\paragraph{Outcome variables}
In output of the experiment we evaluate the accuracy of the assessments made by the experiment subjects. 
We collect analyst output through a survey compiled by all subjects at the end of the experiment.
To evaluate the subject's output under different experimental conditions, we consider the following variables:
    (1) the total number of reports submitted by each group: these may be related to the real injected attacks, or to other suspicious events detected by the subjects;
    (2) the number of submitted reports dealing with one of the simulated attacks;
    (3) the correctness of the submitted reports when compared against the  ground truth.

\subsection{Expected outcome and evaluation criteria}
If the proposed solution is effective in detecting changes in SOC performance triggered by (small) changes in the SOC configuration, we would expect that \GOODSOC{} (i.e., the configuration producing fewer alerts not related to the university's use cases), lead to more accurate reports when compared to \BADSOC{}.
Reflecting normal SOC operation procedures~\cite{muniz2015security}, we split our evaluation over the analysis phases of \textit{Attack identification} and \textit{Attack investigation}.

\smallskip
\textbf{Outcome expectation.}  As the modification between \BADSOC{} and \GOODSOC{} should only reduce the number of false alerts displayed to the analyst, leaving alerts relevant to an attack unaffected, we expect the \SAIBERSOC{} method to highlight that:

\medskip
\noindent (1) \GOODSOC{} outperforms \BADSOC{} for the \textit{attack identification} phase;

\smallskip
\noindent (2) no significant difference between \GOODSOC{} and \BADSOC{} emerges for the \textit{attack investigation} phase. 
    
\smallskip
To evaluate the accuracy of the \textit{Attack identification}, we evaluate the number of correct entries reported by the subjects in relation to: IP(s) of the attacker, IP(s) of the victims. To evaluate the accuracy of \textit{Attack investigation}, we evaluate reported information on the reconnaissance, exploitation, and delivery and control activities (ref. Table~\ref{tab:phases}). 
As we are interested in testing the effectiveness of the proposed solution, we are not concerned with quantifying whether our \GOODSOC\ configuration is \textit{significantly better than} our \BADSOC\ configuration; differently, we are interested in evaluating whether the (albeit small) \textit{change in the \GOODSOC\ configuration can be spotted by the proposed experimental procedure} in terms of a difference in performance. 
To evaluate these differences, we employ a mix of non-parametric statistical tests including Fisher's Exact Test (for differences in counts across conditions), and Wilcoxon rank-sum tests (for differences in outcome distributions). The significance level is set at $\alpha=0.05$.


%

\subsection{Experiment Preparation}

This section details the experiment setup and preparation for the scenario injection, and the attack reporting.

\smallskip
\emph{Attack traces.} We use the previously identified attack phases in Table~\ref{tab:phases} for our proof-of-concept implementation, namely \textit{Reconnaissance}, \textit{Exploitation}, and \textit{Delivery} and \textit{Control} phases. For each of these phases we collect or reconstruct the respective network traces from public resources. For the \textit{Reconnaissance} phase we collect data using \texttt{nmap}\cite{nmap} scans against selected targets;
in the \textit{Exploitation} phase we rely on PoC exploits and brute-force attacks available on public resources, such as ExploitDB\cite{exploitdb} and Metasploit\cite{metasploit}. \textit{Delivery} and \textit{Control} include the download of malicious software from both ExploitDB and suspect IP addresses, and communication with IPs known to be involved in botnet C2C infrastructures.


\smallskip
\emph{Attack generation.} Using the previously identified attack traces, we build two different attack scenarios; these attacks are built by assembling the previous network captures created and provided with consistency among individual actions by editing the IP addresses involved.
Resulting attacks reproduce the \textit{modus operandi} implemented by two real-world attacks (namely, \textit{Mirai} and \textit{Exim}).

\smallskip
\textbf{Scenario 1: Mirai.}
This scenario is inspired by the Mirai botnet\cite{mirai-botnet}.
The attacker gains access by successfully guessing SSH credentials. In the end of the scenario, the victim scans the internal network. As this scenario employs blacklisted IP addresses and domain names related to the Mirai botnet, it raises multiple (high priority) alerts in the \SOC{}.

\textbf{Scenario 2: Exim.}
The second scenario is based on a remote code execution vulnerability in an Exim 4 SMTP mail server \cite{exim-cve}. This scenario raises exactly one (high priority) alert in our \SOC{}.

\smallskip
Table~\ref{tab:scenario_comparison} provides a comparison of the two scenarios across the attack phases defined by MITRE ATT\&CK framework, and which phases can be detected by our \SOC\ configuration. Note that the scenarios are unaffected by the \GOODSOC\ or \BADSOC\ configurations, as the \GOODSOC\ configuration only removes 'policy violation' rulesets that are not triggered by either scenario. Due to the higher number of alerts related to the \MIRAI{} attack than for \EXIM{}, we consider the latter to be a more advanced scenario than the former.


\begin{table*}[t]
    \caption{Comparison between the injected attack scenarios \MIRAI{} and \EXIM{}}
    \label{tab:scenario_comparison}\vspace*{-6pt}
\begin{tabular}{@{}l}
\includegraphics[]{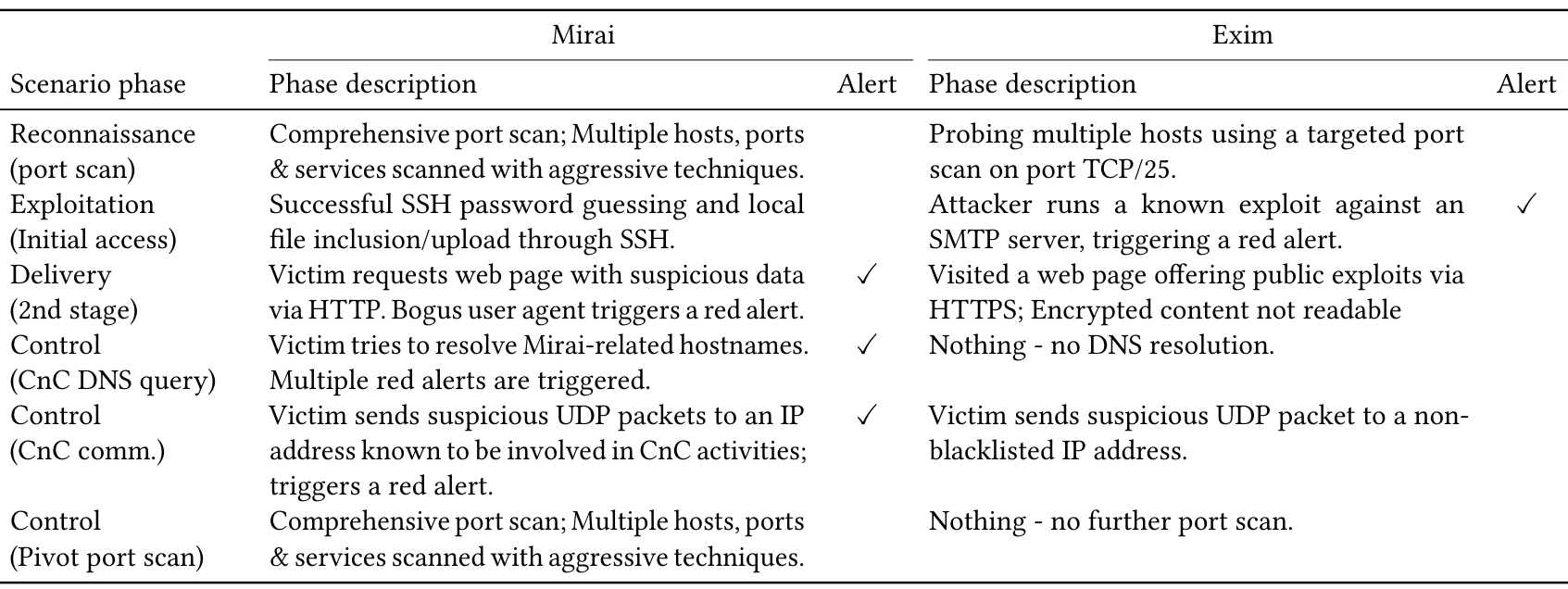}
\end{tabular}
\end{table*}

\smallskip
\emph{Attack injection.} The IP address of the attacker was rewritten to match the IP address defined in the scenarios. The attacks were injected in pre-recorded network traffic arriving at the \SOC\ during working hours. While we cannot guarantee the background traffic to be attack-free, we analyzed the most significant alerts observed in the traffic to ensure that generated alerts not related to the injected attacks are not symptomatic of known attacks. We found no evidence of ongoing attacks in the pre-recorded network traffic.

\smallskip
\emph{Attack report.} Due to infrastructural limitations, we collect reports using an online survey.
\footnote{A \SOC{} would normally rely on an integrated case management system for incident reporting; this was not available in the employed \SOC{} setup.}
Each group could report up to five\footnote{The limit of five activities was chosen after piloting. A discussion of implications on result validity is provided in section~\ref{sec:limitations}.} suspicious activities detected during the experiment.
Table~\ref{tab:questionnaire} provides an overview of the questions.
The first two questions asked for the attacker and victim IP addresses; as multiple entities can be associated to a role for an attack, we allow to insert more than a single IP address. Further, multiple-choice questions address reconnaissance, vulnerability exploitation, the delivery and control phase of the attack.
For each of the selected answers it is required to specify the IP address involved with the selected action.
Lastly subjects could report additional free-text comments. The full survey is available in the Appendix.
To evaluate whether a report is a response to one of the injected attack simulations, we compare the reported IP addresses to the ground-truth from the scenario definition. If the attacker \textit{and} victim IP addresses of a scenario were both mentioned in the respective field, we mark the report to be a response to one of the corresponding scenarios.

\begin{table}[t]
    \centering
    \caption{Summary of information expected in a report}
    \label{tab:questionnaire}
    \Description{Asking for the attacker and victim IP addresses allows to frame the reported scenario, i.e., identify whether the report is in response to one of the injected scenarios or not. Questions about the reconnaissance phase, the exploitation phase, and the delivery and control phases allow to identify whether the respective phase was correctly identified by the SOC analyst. Lastly space for comments allows analysts to provide further information.}
    \begin{tabular}{p{0.3\linewidth} p{0.6\linewidth}}
        \toprule
        {Analysis param}   & {Info provided for result evaluation} \\\midrule
        {Attacker IP}      & Framing of the reported scenario      \\
        {Victim IP}        & Framing of the reported scenario      \\
        {Reconnaissance}   & Detection of reconnaissance actions   \\
        {Exploitation}     & Detection of exploitation actions     \\
        {Delivery \& C\&C} & Detection of delivery \& C\&C         \\
        {Comments}         & Space for further information         \\
        \bottomrule
    \end{tabular}
\end{table}

\smallskip
\emph{Technical setup.} We deployed two instances of the SOC environment and setup each accordingly to the \GOODSOC{} and \BADSOC{} configurations (ref. subsection~\ref{subsec:experimental-variables}). Each VM is powered by a machine with Intel(R) Xeon(R) Bronze 3104 CPU @ 1.70GHz with 12 cores and 64GB of RAM.
During the experiment the two \SOC{} servers' sensors were isolated from the University's network. Students accessed their respective analysis environments from the classroom on their own PCs.

%
\subsection{Experiment Pilot}

Prior to execution we ran a set of pilots to refine the infrastructure and test the devised experimental procedures. A pilot phase was conducted to evaluate the survey procedure (incl. question phrasing) employed for the reporting phase. Further, the pilots were used to assess whether the difficulty of the assignment and the pre-experiment training sessions were a good match with students' skills. To this aim, we piloted the whole experiment four times, pooling for volunteers from students with a similar background to our subjects (e.g., enrolled in other security courses at the time, or PhD students in Computer Science).

\subsection{Experimental Execution}

Student autonomously created groups of two and each group was assigned randomly to the \BADSOC\ (n=31) or \GOODSOC\ (n=32) experimental conditions. Students were given half a bonus point valid for the final exam for attending the experiment and if they managed to correctly identify (at least one) attacker in the injected attacks.

\begin{figure}[t]  
 \centering
    \includegraphics[width=0.5\columnwidth]{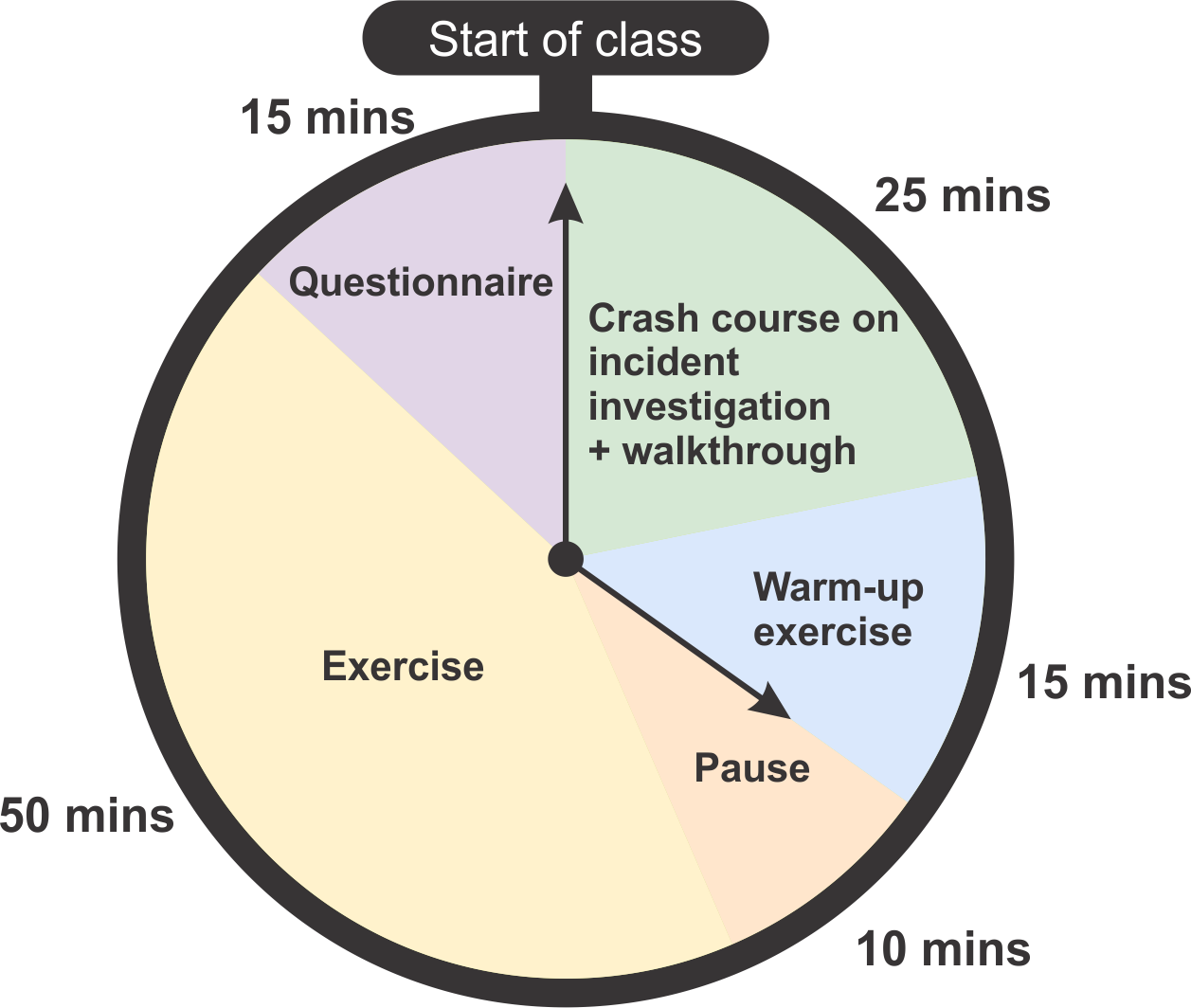}
    \caption{Experiment phases and duration}
    \label{fig:exptimeline}
    \Description[A timeline describing the individual phases of the experiment session and their duration.]{The experiment session lasted 115 minutes. First: 25 minutes crash course on incident investigation and a (SOC) walk-trough. Then: 15 minutes warm-up exercise. Then: 10 minutes break. Then: 50 minutes exercise (main experiment). Then: 15 minutes questionnaire.}
\end{figure}

The experiment session started with an introduction and training session lasting approximately 30 minutes. After introducing the assignment, we ran a 15 minutes warm-up exercise where students had to identify a simple scenario with small network traffic to familiarize with the interface. The warm-up exercise was run on the \BADSOC\ configuration for all groups.\footnote{The warm-up exercise, albeit simpler and shorter than the actual exercise, follows the same structure as the subsequently injected attacks.}
The experiment was run after the warm-up phase and lasted 50 minutes.
Subjects were told that at least one attack would be injected but were unaware of attack type and timing.
At the end of the experiment, traffic injection was terminated, and the survey used for report collection was made available.
Students were asked to submit their reports within 15 minutes but retained access to \SOC{} and questionnaire until all submissions were collected.
Figure~\ref{fig:exptimeline} provides a visual representation of the experiment timeline.

%
\subsection{Ethical and experimental considerations}
\label{sec:limitations}

The background noise traffic was captured from the \SOC\ operating at the M\&CS department of the host university under the ethical and organizational approval of the university. The traffic raw data was not accessible to the students during the experiment (i.e., they could only see the alerts generated based on the original traffic). The experiment was integrated in the practical educational activities run by the university and performed following guidelines approved by the university for exercises in classroom settings.

\smallskip
\emph{Experimental limitations.} \textbf{(a) Construct validity.}
We implicitly assume that students who report the correct IP addresses for both attacker and victim detected the attack and started analyzing it.
Using automated checking it is impossible for us to determine whether a group actually understood what happened.
\textbf{(b) External validity.}
Students do not have the same experience and training that \SOC{} analysts do and the 
training during the introduction section cannot make up for that. 
In real settings, the experiment should be replicated against multiple analysts and using a portfolio of attacks, as opposed to only one or two attack scenarios, to obtain ample validity across experimental conditions.
\textbf{(c) Internal validity.}
Students were told in advance that there would be at least one attack, creating an expectation to find something. 
Further, by allowing students to submit up to five incidents and not penalizing `false positive' reporting, our experiment may inflate the number of reports per group. As this limit is the same across all groups and treatments, we do not expect any effect on the likelihood that groups report the `real' injected incidents (i.e. our main experimental outcome).

%
\section{Experimental Results}
\label{sec:experimentresults}

Whereas students could submit up to five reports, we expected exactly two reports detecting the injected attack scenarios \MIRAI{} and \EXIM{} respectively.
Table~\ref{tab:expectations} reports the correct answers related to the \MIRAI\ and \EXIM\ scenarios.

\begin{table}[t]
    \centering
    \caption{Expected findings/ground-truth for both scenarios}
    \label{tab:expectations}
    \Description{We expect reports in response to the Mirai attack to mention \MIRAIATTACKERIP{} as attacker IP address, \MIRAIVICTIMIP{} as victim IP address, 'port scan` as reconnaissance action, `weak credentials' as vulnerability or exploit used, and `data exfiltration' and `HTTP requests' as activities observed in the Delivery and Command and Control phase of the attack. In response to the Exim attack we expect reports to mention \EXIMATTACKERIP{} as attacker IP address, \EXIMVICTIMIP{} as victim IP address, 'port scan` as reconnaissance action, `remote code execution' as vulnerability or exploit used, and `data exfiltration' and `HTTP requests' as activities observed in the Delivery and Command and Control phase of the attack.}
    \begin{tabular}{p{0.26\columnwidth} p{0.30\columnwidth} p{0.30\columnwidth}}
        \toprule
        {Analysis param} & \MIRAI                                                                     & \EXIM{}                    \\\midrule
        {Attacker IP }   & \MIRAIATTACKERIP{}                                                         & \EXIMATTACKERIP{}          \\
        {Victim IP }     & \BLINDED{\MIRAIVICTIMIP{}}                                                 & \BLINDED{\EXIMVICTIMIP{}}  \\
        {Reconnaissance} & port scan                                                                  & port scan                  \\
        {Exploit/Vuln.}  & weak credentials                                                           & remote code exec.          \\
        {Delivery/C\&C}  & data exfiltration\hfill\break HTTP requests  & data exfiltration\hfill\break HTTP requests\\
        \bottomrule
    \end{tabular}

\end{table}

\subsection{Attack identification}

\begin{figure}[t]
    \centering
    \includegraphics[width=\linewidth,height=6cm,keepaspectratio]{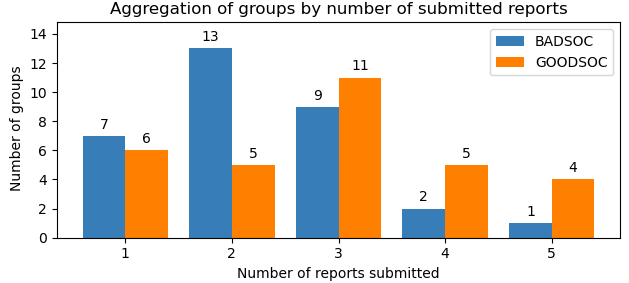}
    \caption{Aggregation of groups by \# submitted reports}
    \label{fig:plot-submission-histogram}
    \Description{Participants were allowed to submit up to five potential attacks. In the BADSOC condition, 7 groups submitted one report, 13 groups submitted two reports, 9 groups submitted three reports, 2 groups submitted four reports, and 1 group submitted five reports. In the GOODSOC condition, 6 groups subitted one report, 5 groups ubsmitted two reports, 11 groups submitted three reports, 5 groups submitted four reports, and 4 groups submitted five reports.}
\end{figure}

We received 63 submissions (one per group) reporting a total of 162 incidents. 
Figure~\ref{fig:plot-submission-histogram}
reports overall submission rates for all groups in the \BADSOC{} ($m=2.28,\ sd=0.99$) and \GOODSOC{} ($m=2.9,\ sd=1.3$) configurations.
Overall, the 32 groups working on \BADSOC{} submitted 73 reports, the 31 groups on \GOODSOC{} submitted 89 reports in total.
%
%
A Wilcoxon Rank-Sum test results in \BADSOC{} producing significantly less reports than \GOODSOC{} ($p=0.02,\ W=353.5$). 
This suggests that groups assigned to the \GOODSOC{} condition were able to reconstruct more `suspicious' activities than groups assigned to \BADSOC{}, despite the events related to these activities being available to both treatment groups.

\begin{figure}[t]
    \centering
    \includegraphics[width=\linewidth,height=6cm,keepaspectratio]{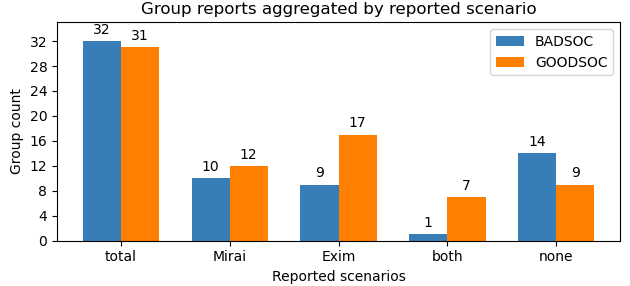}
    \begin{minipage}{0.9\columnwidth}
        \footnotesize
        \smallskip
        Column \enquote{both} represents the number of groups that reported both scenarios (\MIRAI{} \emph{and} \EXIM{}), while \enquote{none} represents the number of groups that did not report any of the scenarios.
    \end{minipage}
    \caption{\textit{Attack identification} by reported scenario.}
    \label{fig:plot-group-report-count}
    \Description{In total, 32 groups operated in the BADSOC configuration, 31 groups in the GOODSOC condition. In BADSOC 14 (out of 32) groups did not report any attack, 10 reported the Mirai and 9 the Exim attack. Out of the 18 groups that found at least one attack, one reported both attacks. In GOODSOC 9 (out of 31) groups did not report any attack, 12 reported the Mirai and 17 the Exim attack. Out of the 22 groups that found at least one attack, 7 reported both attacks.}
\end{figure}

Overall, $26\%$ of \BADSOC{} reports (19/73) and $33\%$ of \GOODSOC{} reports (29/89) detailed the injected attack scenarios (i.e., the majority of reports were \textit{not} about the injected attacks).
As groups could report up to five incidents, this is unsurprising.
To evaluate reporting in more detail, Figure~\ref{fig:plot-group-report-count} 
shows the distribution of reports on the injected scenarios. 
Overall, 57\% of groups in \BADSOC{} and 71\% of groups in \GOODSOC{} reported at least one of the injected attacks.
Whereas no noticeable difference against \MIRAI{} reporting rates can be observed ($p=0.36,\ OR=1.38$), \GOODSOC{} groups were approximately three times more likely to report the \EXIM{} attack ($p=0.03,\ OR=3.04$) than the \BADSOC\ groups, and approximately eight times more likely to report both attacks ($p=0.02,\ OR=8.49$) overall.
However, if one considers only groups that reported at least one of the attacks, the difference between \GOODSOC{} and \BADSOC{} decreases and becomes borderline significant ($p=0.09$,\ $OR=5.56$);
this may suggest that, under our treatment conditions, groups that perform well and detect at least one attack are more likely to find both irrespective of the \SOC\ configuration.
Similarly, \BADSOC\ groups seem overall more likely ($OR=1.88$) than \GOODSOC{} to \textit{not} report either of the attack scenarios, albeit the difference is not significant ($p=0.17$).

These findings provide an initial indication that the proposed solution can be effectively employed to evaluate (relative) SOC performance across SOC configurations (e.g., before and after a major configuration change in a deployed \SOC{}).


\subsection{Attack investigation}

We now provide a breakdown over the \emph{reconnaissance}, \emph{exploitation}, and \emph{delivery \& control} attack phases across the two injected attack scenarios. Results are reported in Table~\ref{tab:content-result-table}.



\begin{table*}[t!]
\centering
\caption{Report results of \textit{attack investigation}.}
\label{tab:content-result-table}\vspace*{-6pt}
\begin{tabular}{@{}l}
\includegraphics[]{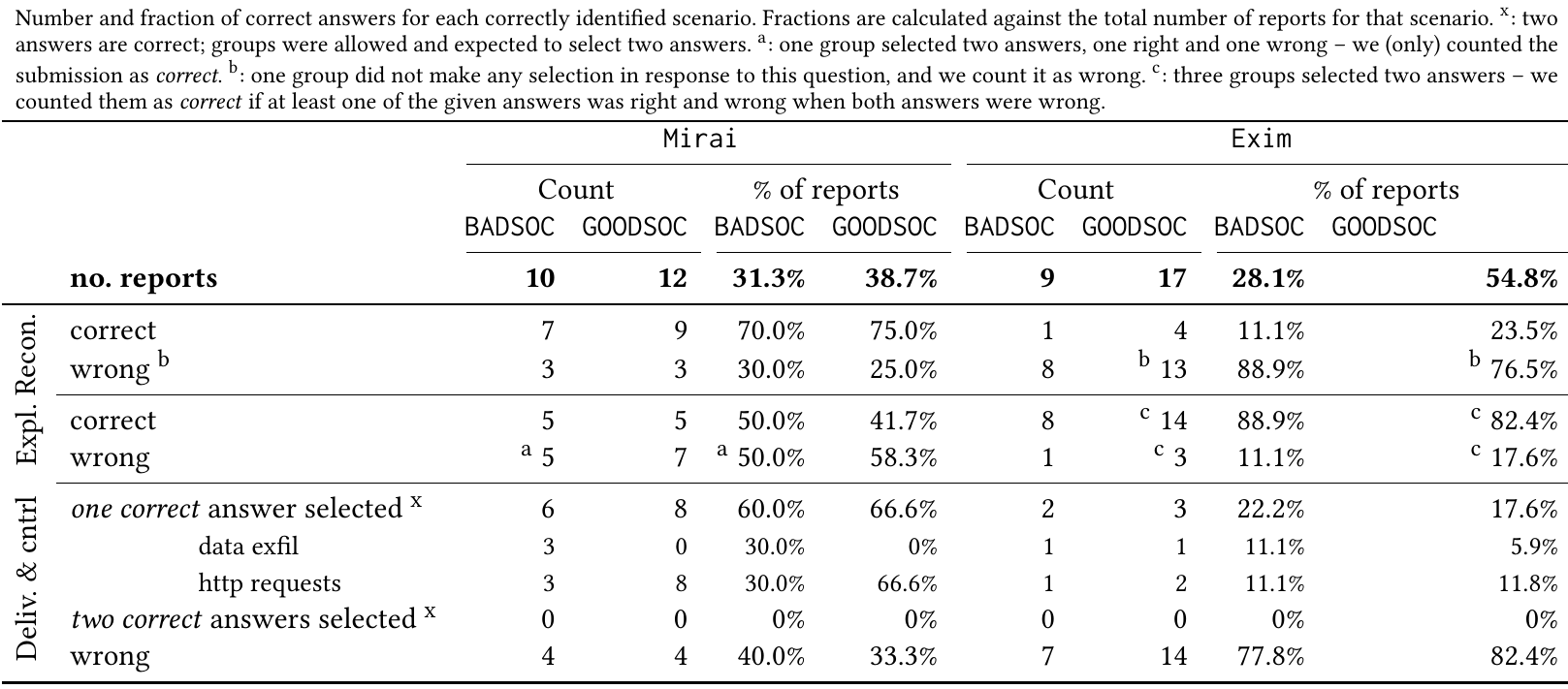}
\end{tabular}
\end{table*}



%

\paragraph{Reconnaissance.}
For both injected attacks, \enquote{port scan} was the correct answer, when asked for the observed reconnaissance activity in the questionnaire (see Table~\ref{tab:expectations}).
For the \MIRAI{} attack, the correct answer was given by 70\% (7/10) groups operating on \BADSOC{} and 75\% (9/12) groups operating on \GOODSOC{} configuration ($p=0.58,\ OR=1.27$).
The ratio of reports mentioning the portscan for the \EXIM{} attack is, overall, lower than for \MIRAI{}, albeit no significant difference emerges ($p=0.42,\ OR=2.39$):
In \BADSOC{} 11.1\% (1/9) correct answers, in \GOODSOC{} 23.5\% (4/17). 
In conclusion, \GOODSOC{} did not perform significantly better than \BADSOC{} with in the detection of the portscan in none of the scenarios.

%


\paragraph{Exploitation.}
For exploitation activity, Table~\ref{tab:content-result-table} shows that barely half of the groups selected the correct answer \enquote{weak credentials} when asked to report the exploitation vector abused by the attacker in the \MIRAI{} attack, i.e. 50\% (5/10) in \BADSOC{} and 41.7\% (5/12) in \GOODSOC{}.
For the \EXIM{} attack, Table~\ref{tab:content-result-table} shows that 88.9\% (8/9) of the groups in \BADSOC{} that reported the \EXIM{} attack also reported the correct vulnerability \enquote{remote code execution} abused by the attacker to gain foothold on the victim.
In \GOODSOC{}, 82.4\% (14/17) answered correctly.
On neither of the two injected attacks \GOODSOC{} performed better than groups working with the \BADSOC{} (\MIRAI{}: $p=0.79,\ OR=0.73$, \EXIM{}: $p=0.84,\ OR=0.59$).
In fact, \BADSOC{} performed slightly better than \GOODSOC{} in identifying the correct exploitation vector (\MIRAI{}: $p=0.52,\ OR=1.38$, \EXIM{}: $p=0.57,\ OR=1.68$).
Putting the differences between \BADSOC{} and \GOODSOC{} aside, all groups were more likely to identify the correct exploitation actions in \EXIM{} (85\%, or 22/26) than in the \MIRAI{} scenario (45\%, or 10/22).

%

%

\paragraph{Delivery \& Control.}
We expected groups to select two correct answers in a multiple-choice question about the delivery and control activities observed in context of an attack.
The answers \enquote{data exfiltration} and \enquote{HTTP requests} are considered correct for both simulated attacks (see Table~\ref{tab:expectations}).
Of the groups who reported on the \MIRAI{} attack, 66.6\% (8/12) using \GOODSOC{} and 60\% (6/10) of the groups using \BADSOC{} reported at least one of the two activities ($ p=0.55,\ OR=1.32 $).
Looking at the \EXIM{} attack, 22\% (2/9) of the groups reporting the attack on \BADSOC{} and 18\% (3/17) on \GOODSOC{} also mentioned one correct delivery/control activity ($p=0.79,\ OR=0.76$).
No group submitted both correct answers, regardless of their condition or reported attack.
Again, there are no significant differences in performance between \BADSOC{} and \GOODSOC{}.

In the \MIRAI{} scenario option \enquote{HTTP requests} was likely to be selected roughly four times more often ($p=0.099,\ OR=4.32$) by groups operating \GOODSOC{}.
In the \BADSOC{} configuration 30\% (3/10) selected it, in \GOODSOC{} 67\% (8/12).
Only three groups, all using \BADSOC{}, chose the correct answer \enquote{data exfiltration} when reporting the \MIRAI{} attack.
For the \EXIM{} attack, most groups, 56\% (5/9) in \BADSOC{} and 71\% (12/17) in \GOODSOC{} \emph{wrongly} chose \enquote{none of them}.

%
\section{Discussion of results and presentation of the \SAIBERSOC{} tool}
\label{sec:discussion}

\begin{figure*}[t]
\includegraphics[]{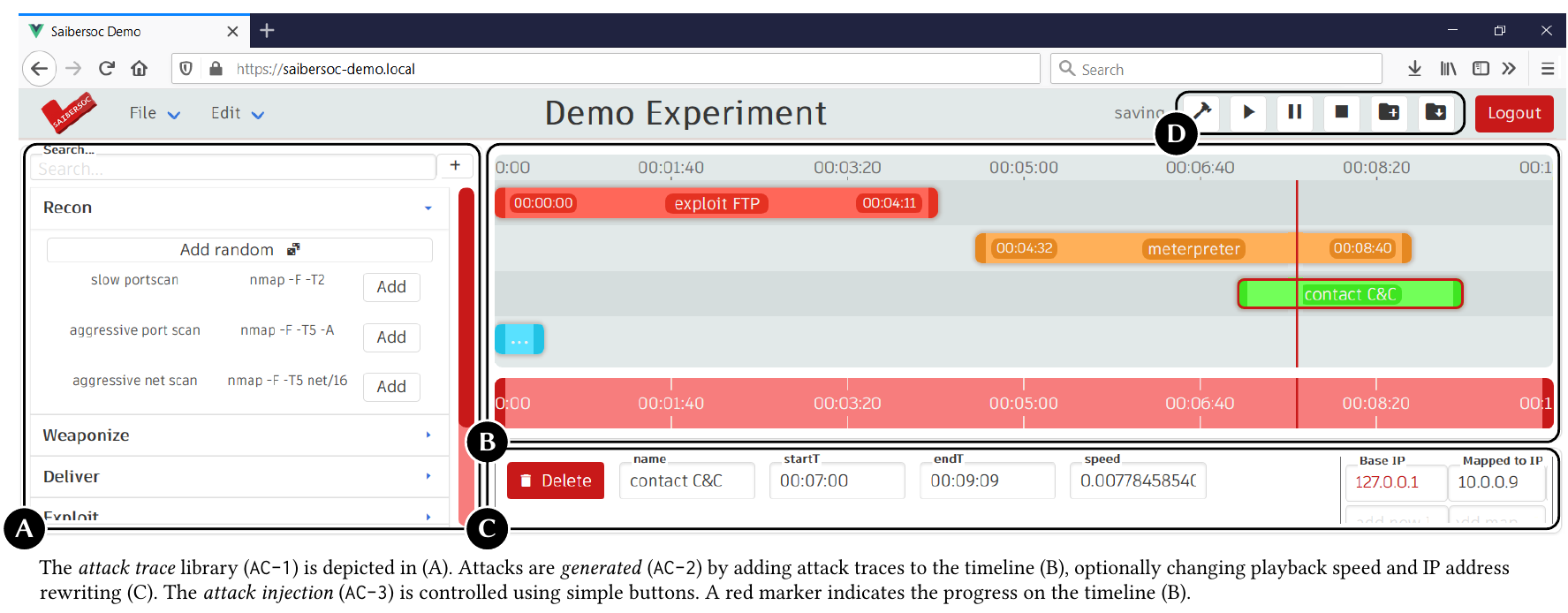}   
 \caption{Screenshot of the \SAIBERSOC{} web interface and relation to the relevant architectural components.}
    \label{fig:saibersoc-ui}
    \Description[A screenshot of the SAIBERSOC user interface.]{\SAIBERSOC{} allows to generate attacks based on attack traces and inject them into the network. The picture shows three different views: the attack-trace library, the view to generate a new experiment, and lastly a timeline view  of an assembled attack that can be injected at a later point in time.}
\end{figure*}

The presented results indicate that the \SAIBERSOC{} methodology is capable of highlighting the modifications introduced between SOC configurations. In our experiment, the \GOODSOC{} configuration significantly increases the chances of a correct \textit{attack identification}; however, the \BADSOC{} vs \GOODSOC{} conditions do not significantly affect the chances of correctly investigating the attack across its phases, indicating that the change in configuration leaves \textit{attack investigation} capabilities unaffected. 

Overall, in our example application the \SAIBERSOC{} method indicates that the \GOODSOC\ configuration is preferable to the \BADSOC\ configuration as it significantly increases (up to threefold) the likelihood of identification of incidents. 
In particular, if \GOODSOC\ and \BADSOC\ were alternative environments tested in a SOC environment, using the \SAIBERSOC\ solution the SOC manager of that infrastructure could conclude that analysts working with the \GOODSOC{} configuration were more likely to find one or both of the attacks: analysts on \GOODSOC{} were three times more likely to find the (slightly harder to identify) \EXIM{} attack ($p=0.03,\ OR=3.04$) than analysts on \BADSOC.
The method results indicate that analysts in the \BADSOC{} configuration are either overwhelmed by the number of alerts to investigate or require more time analyzing individual events and therefore had no time to analyze the second scenario.
However, analysts working with \GOODSOC{} produced more false-positive reports, i.e., reports \emph{not} in response to one of the injected attacks. This may indicate that analysts working on \GOODSOC{} either spent less time per analysis and thus could write more reports in the same time, or that they are generally more likely to report false-positives.
\footnote{This effect may be caused by our experiment design; students had the \textit{expectation} to find at least one attack and were rewarded on finding at least one of the injected attacks, which may lead to over-reporting.}
By contrast, no significant differences between \BADSOC{} and \GOODSOC{} can be observed in the \textit{attack investigation} phase. This provides additional information on the SOC performance related to which aspects of an incident investigations are (positively or negatively) affected by the different configurations under test. The ability to quantify these effects experimentally allows both researchers and practitioners in making informed decisions on technological and process-level solutions for security monitoring application.
\smallskip
\emph{Implications for research.}
The \SAIBERSOC{} method can be used to evaluate the effect on security analysis of virtually any live analysis setting (such as a SOC or other security monitoring or analysis environments), either technical, procedural, or human.
For example, \SAIBERSOC{} can be employed to evaluate the impact of factors such as the analysts' experience, skill composition, and training, or the effect of analysis and visualization tools by comparing the performance of two \SOC{} configurations with a known set of selected differences.
Similarly, the proposed method can be employed to evaluate multiple performance variables, including timing (e.g., total time for alert investigation) and throughput measures~\cite{shah2018methodology}), and how these vary across different phases of the attack.
The extensibility of the proposed method further opens the door to the investigation of more complex effects, such as the relation between attack types and analyst skills/expertise~\cite{emse-20}, which are still under-researched in the scientific literature.
The \SAIBERSOC{} implementation presented further down in this section provides a practical tool to deploy experiments based on the propose method in live or laboratory environments.

\smallskip
\emph{Implications for practice.}
The \SAIBERSOC{} methodology (and tool) is fully flexible in terms of attack procedure, traces, and outcome variables, and can be implemented in operational \SOC{}s to evaluate the relative performance of their configuration over the desired metrics.
This includes both technical (e.g., rulesets, use-cases, alert correlation rules) and human-level (e.g., analysis processes, analyst expertise, flat vs. hierarchical reporting structures) aspects of SOC operation, both of central importance for effective incident detection and reporting~\cite{kokulu2019matched,muniz2015security}.
Examples of concrete use-cases are comparison of \SOC{} performance before and after changes in \SOC{} configuration, or continuous monitoring of \SOC{} performance.
Further, the proposed solution can be integrated in analyst training procedures, or as part of a selection procedure for security professional roles.

%
\subsection{\SAIBERSOC\ tool implementation}
\label{subsec:saibersoc-tool}

To operationalize the \SAIBERSOC{} methodology, after its validation, we developed the \SAIBERSOC{} tool as an open source software solution.
\SAIBERSOC{} implements all four architectural components outlined in subsection~\ref{subsec:architecture}. It is deployed with a web-based front-end (depicted in Figure~\ref{fig:saibersoc-ui}) interacting with a backend. A comprehensive API enables scripted access to all backend functionalities.

\emph{Functionalities and User Interface.}
Figure~\ref{fig:saibersoc-ui} shows the deployed web interface and how architectural components (see subsection~\ref{subsec:architecture}) \arch{1}\ldots\arch{3} are implemented in the web interface; report evaluation (\arch{4}) is not yet accessible through the frontend. 
On the left side of the screen, UX element \encircle{A} provides access to the attack \emph{trace library} (\arch{1}) showing previously uploaded network traffic recordings.
A user can manually add traces onto a timeline \encircle{B} in the center of the screen.
Convenience features allow to search a specific attack trace in the library or select a random one.
From the timeline \encircle{B}, the offset and playback speed can be adjusted for each attack trace individually.
In the lower section of the screen, \encircle{C} provides additional configuration options for the selected attack trace. Here, IP addresses and whole IP address ranges can be rewritten to create a realistic attack scenario for the monitoring environment.
Buttons \encircle{D}, in the upper right area, provide functionalities to load and save experiments/attacks and to start, stop, or schedule their injections (\arch{3}).
During attack injection (\arch{3}), state and progress is visualized on the timeline \encircle{B} using a vertical red line as progress indicator.
The tool is capable of automatically extracting a ground-truth based on the metadata from the attack trace library (\arch{1}) and the optional IP address rewrites in \arch{2}.
A \SOC{} report, allegorized by a \texttt{csv} file, can be checked against the ground-truth derived from the attack traces (\arch{4}).

\emph{Tool implementation and API.}
The \SAIBERSOC{} web application is implemented in \texttt{Python}.
The (HTTP) API is based on \texttt{Flask}\cite{flask} and the \texttt{Flask-RESTX} framework\cite{flask-restx}.
The web-frontend is built on top of the API using \texttt{vue.js}\cite{vuejs}.

\noindent\textbf{Attack Traces~(\arch{1})}:
The attack trace library consists of attack traces in form of \texttt{pcap} network traffic recordings linked to metadata stored in an internal database.\hfill\break
\noindent\textbf{Attack Generation~(\arch{2})}:
Internally, \SAIBERSOC{} uses a list of \emph{blocks} to specify an attack. Each \emph{block} consists of an attack trace (taken from the attack trace library) and additional information specifying time offset, playback speed, and IP address rewriting.
Once an attack is fully sketched out, \texttt{scapy}\cite{scapy} is used to assemble a single intermediate attack \texttt{pcap}.\hfill\break
\noindent\textbf{Attack Injection~(\arch{3})}:
\SAIBERSOC{} invokes \texttt{tcpreplay}\cite{tcpreplay} to inject the \enquote{intermediate attack \texttt{pcap}} on a network interface.
Optionally, background noise can be replayed together with the attack.\hfill\break
\noindent\textbf{Reporting~(\arch{4})}:
The reporting module extracts and combines the IP address rewriting from \arch{2} and attack trace metadata from \arch{1} as ground-truth.
It then compares selected columns from an uploaded \texttt{csv} file against the extracted ground truth.

\emph{Publication, Development, and Licensing}
The development of \SAIBERSOC{} was supported by a team of BSc students as part of their final graduation project.
The \SAIBERSOC{} tool is released under Mozilla Public License (MPL 2.0).
Source code, extensive documentation, and supplementary materials are available from the artifact repository at \url{https://gitlab.tue.nl/saibersoc/acsac2020-artifacts}.%


%
\section{Conclusion}
\label{sec:conclusion}

In this paper, we proposed a methodology based on attack injection to systematically measure \SOC{} performance across SOC configurations, analyst expertise, and for any output metric (e.g., accuracy, time-to-report, \ldots).
We verified the proposed methodology by conducting an experiment in which 124 students assumed the role of a \SOC{} analysts.
Our results show that the proposed methodology is capable of systematically measuring \SOC{} performance and attributing it to differences in configuration, where some is to be expected.
Our solution is general and can be implemented in any SOC.
In addition, we developed the \SAIBERSOC{} tool to help replicate the experiments and facilitate SOC exercises.


\hyphenation{DEFRAUDIfy}
\begin{acks}
    This work is supported by the DEFRAUDIfy project (grant no. ITEA191010) funded by the ITEA3 program by Rijksdienst voor Ondernemend Nederland and the DEPICT project (grant no. 628.001.032) by Nederlandse Organisatie voor Wetenschappelijk Onderzoek.
\end{acks}

\balance
\bibliographystyle{ACM-Reference-Format}
\bibliography{arxiv}


\appendix
\balance


\section*{Appendix}

\begin{table*}[t]
    \centering
        \caption{Mapping between requirements and architectural components.}
    \label{tab:requirements-implementation-mapping}
    \begin{tabular}{l p{0.21 \textwidth} p{0.21\textwidth} p{0.21\textwidth} p{0.21\textwidth}}
        \toprule
        \require{ID}  & \arch{1} & \arch{2} & \arch{3} & \arch{4} \\
        \midrule
        \require{1}     & Attack traces are independent of the monitored environment.  & The attack generation allows for the matching of the traces to the environment. &      & The reporting and comparison relies on the ground truth for the generated scenario. \\
        \require{2}      & The attack traces identify all MITRE ATT\&CK `building blocks'. & The generated attack can suit any final environment and attack scenario. &      &      \\
        \require{3}          &      & Attacks can be mapped to the target environment using asset information always available at the SOC. & The injection happens in the real network flow monitored by the SOC.  &      \\
        \require{4}    &      & The tailoring of the attack generation guarantees the existence of a ground-truth for the examination.  &      & The analyst report can be automatically checked against the attack parameters (\arch{2}). \\
        \bottomrule
    \end{tabular}
\end{table*}

\begin{table*}[t]
    \centering
    \caption{Implementation example across \arch{1}\ldots\arch{4}.} 
    \label{tab:example}
\begin{tabular}{@{}l}
\includegraphics[]{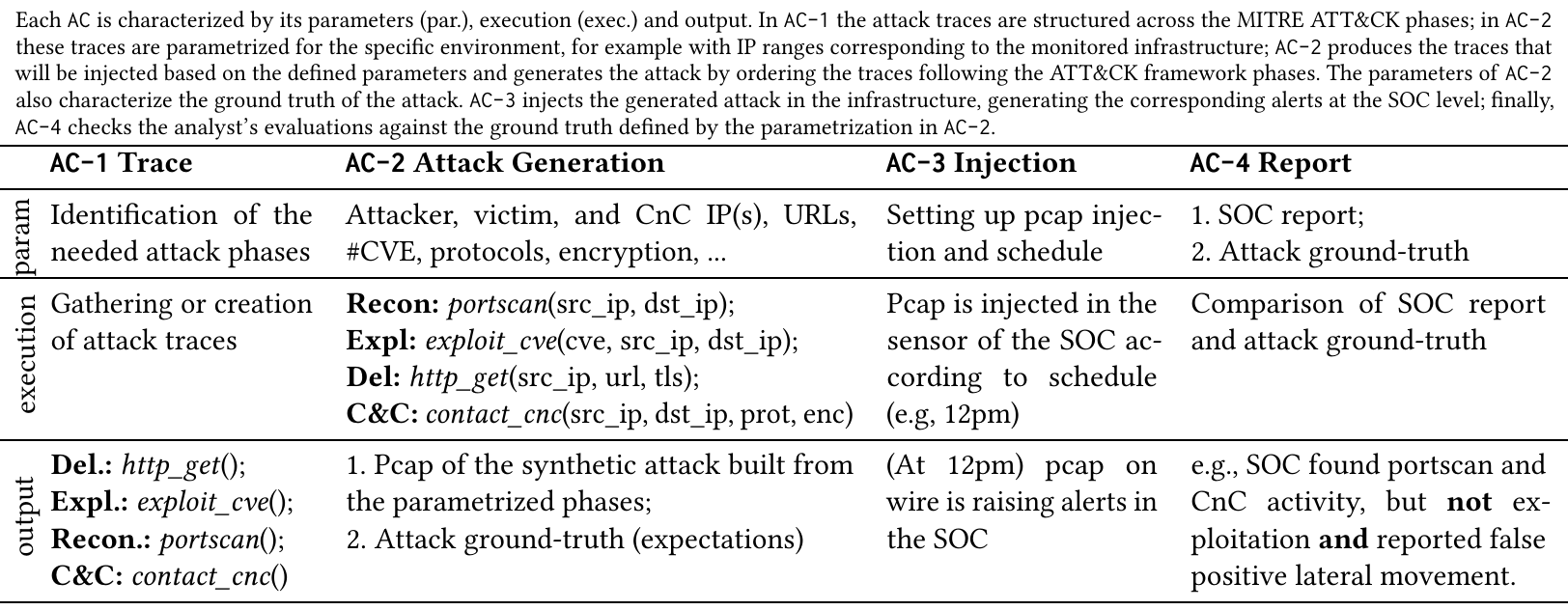}
\end{tabular}
\end{table*}

\subsection*{Data Collection and Analysis}

\paragraph{Evaluation for possible typos}

In an exploratory data analysis, we observed that some groups reported IP addresses that are very similar to the ones we expected to see as attacker or victim IP address in one of the two scenarios.
Based on the difference between the IP addresses in the report and the ground-truth, we conclude that it is very likely that those are typos or copy-paste errors.

In other cases, groups did not make the right selections in the multiple-choice part of the questionnaire, but later mentioned correct details in the additional comment section of the report.
For example, group 78 did not indicate that they observed \enquote{http requests} through the multiple-choice selection but wrote \enquote{[\ldots] this malware made a HTTP request [\ldots]} in the additional comments.

To ensure automatic and reproducible processing of the results, we decided to only consider IP addresses that are perfect matches and solely rely on the multiple-choice selection.

\paragraph{Submissions violating the enforced format}
Some groups managed to submit reports that violate the constraints enforced through the questionnaire.
This mostly includes submissions that ticked more or less checkboxes than we allowed.
Group 80 was able to select three options where the questionnaire was supposed to allow no more than two selections.
We kept those submissions as they do not interfere with our way of data analysis.

\paragraph{Double-submission}
One group submitted twice, i.e. we collected 64 submissions for 63 groups. While both submissions are similar, they are not identical.
We decided to merge the submissions.
The reports in response to one of the injected attack scenarios are merged by selecting the best parts of both submissions.
We further decided to keep the submission time and duration of their first submission.
\paragraph{(Not) Accepting \enquote{lateral movement} in \MIRAI{}}
We initially did not plan to accept \enquote{lateral movement} as a correct answer in \MIRAI{} scenario.
In the \MIRAI{} scenario, the victim eventually mirrors the attacker's behavior and starts scanning the local network for vulnerable hosts.
While the scenario ended before and therefore does not cover lateral movement, the network scan can be interpreted as a first step of lateral movement.
We decided to \emph{not} accept \enquote{lateral movement} as a correct answer for the \MIRAI{} scenario.

In total 40.0\% groups (4/10) in \BADSOC{} and 41.7\% groups (5/12) in \GOODSOC{} selected \enquote{lateral movement} (either alone or in combination with \enquote{http requests}).

\subsection*{Questionnaire}

Based on the choice on question~9, questions~3--8 can be repeated up to four additional times (i.e., a total of five reports can be submitted).
The decision to limit the number of reports per submission was induced by limitations of the survey tool used to collect the results.

\begin{enumerate}
    \item \textbf{Insert your student ID}\hfill\break
          Multiple short text fields for multiple student IDs. Answers are considered personal data.
    \item \textbf{Which was your SOC name?}\hfill\break
          Binary choice between \BADSOC{} and \GOODSOC{} (names changed).
    \item \textbf{Insert the IP address(es) of the attacker, one per line. If no IP address is known, write \enquote{NA}.}\hfill\break
          Multi-line text field.
    \item \textbf{Insert the IP address(es) of the attacker, one per line. If no IP address is known, write \enquote{NA}.}\hfill\break
          Multi-line text field.
    \item \textbf{Did you observe any of the following reconnaissance activities?}\hfill\break
          Multiple choice single selection only: \enquote{Username enumeration}, \enquote{Port scan}, \enquote{SIP scan}, \enquote{Web application vulnerability scan}, \enquote{None of them}, \enquote{Other (small text field to specify)}.
    \item \textbf{Which vulnerability did the attacker exploit?}\hfill\break
          Multiple choice, multiple selections possible: \enquote{SQL injection}, \enquote{Weak credential}, \enquote{DNS remote command execution}, \enquote{Poor web server configuration}, \enquote{Remote coded execution}, \enquote{None of them}, \enquote{Other (small text field to specify)}.
    \item \textbf{Which of the following actions did you observe? (check at most 2 boxes). For each option selected insert IP address of the receiver of such activities (i.e. HTTP requests -> web server's IP; enumerating SMB shares -> SMB server's IP;.}\hfill\break
          Multiple choice, at least one, at most two answers can be selected. Except of \enquote{No action observed}, all answers have a small text field to further specify an IP address: \enquote{Data exfiltration}, \enquote{Enumerating SMB shares}, \enquote{HTTP requests}, \enquote{Denial of Service attack}, \enquote{Web server path traversal}, \enquote{NTP amplification}, \enquote{Network lateral movement}, \enquote{No action observed}.
    \item \textbf{Anything else to report about this attack?}\hfill\break
          Multi-line text field.
    \item \textbf{Do you want to report another attack?}\hfill\break
            Binary yes/no selector. Subjects could submit up to five reports (questions 3 -- 8), selecting \enquote{yes} would allow them to continue, selecting \enquote{no} would allow them to submit their results.
    \item \textbf{Did you had fun?}\hfill\break
            Rating-scale 1 (really boring) -- 5 (it was great)
    \item \textbf{Do you think the introduction was enough to do the exercise?}\hfill\break
            Rating-scale 1 (not at all) -- 5 (definitely enough)
    \item \textbf{Do you think we should do again this next year?}
            Binary yes/no selection.
    \item \textbf{Did you find anything particularly challenging during the investigation process?}\hfill\break
            Multi-line text field.
    \item \textbf{Write here any other suggestions you might have in mind.}\hfill\break
            Multi-line text field.
\end{enumerate}



\end{document}